\RequirePackage{fix-cm}
\documentclass[smallextended]{svjour3}       
\smartqed  
\usepackage{graphicx}
\begin{document}

\title{Terahertz Dynamics of Quantum-Confined Electrons in Carbon Nanomaterials
}


\author{Lei Ren, Qi Zhang, S\'{e}bastien Nanot, Iwao Kawayama, Masayoshi Tonouchi, and Junichiro Kono}

\authorrunning{L. Ren, Q. Zhang, S. Nanot, I. Kawayama, M. Tonouchi, and J. Kono} 

\institute{L. Ren, Q. Zhang, S. Nanot, and J. Kono \at
              Department of Electrical \& Computer Engineering, Department of Physics \& Astronomy,\\
              and The Richard E. Smalley Institute for Nanoscale Science and Technology,\\
              Rice University, Houston, Texas 77005, U.S.A.\\
              Tel.: +1-713-348-2209,
              Fax: +1-713-348-5686\\
              \email{kono@rice.edu}            \\
%
\\
I. Kawayama and M. Tonouchi \at
              Institute of Laser Engineering, Osaka University,
              Yamadaoka 2-6, Suita,\\ Osaka 565-0871, Japan\\
              Tel.: +81-70-6507-9589,
              Fax: +81-6-6879-7984\\
}

\date{Received: date / Accepted: date}

\maketitle

\begin{abstract}
Low-dimensional carbon nanostructures, such as single-wall carbon nanotubes (SWCNTs) and graphene, offer new opportunities for terahertz science and technology.  Being zero-gap systems with a linear, photon-like energy dispersion, metallic SWCNTs and graphene exhibit a variety of extraordinary properties.  Their DC and linear electrical properties have been extensively studied in the last decade, but their unusual finite-frequency, nonlinear, and/or non-equilibrium properties are largely unexplored, although they are predicted to be useful for new terahertz device applications.  Terahertz dynamic conductivity measurements allow us to probe the dynamics of such photon-like electrons, or massless Dirac fermions.  Here, we use terahertz time-domain spectroscopy and Fourier transform infrared spectroscopy to investigate terahertz conductivities of one-dimensional and two-dimensional electrons, respectively, in films of highly aligned SWCNTs and gated large-area graphene.  In SWCNTs, we observe extremely anisotropic terahertz conductivities, promising for terahertz polarizer applications.  In graphene, we demonstrate that terahertz and infrared properties sensitively change with the Fermi energy, which can be controlled by electrical gating and thermal annealing.

\end{abstract}


\section{Introduction}
\label{intro}


The discoveries of C$_{60}$ fullerene by Kroto and co-workers in 1985 \cite{SmalleyNature85} and carbon nanotubes (CNTs) by Iijima in 1991 \cite{Iijima91Nature}, following a preliminary carbon filamentous growth work \cite{EndoJCG76}, have opened up a new scientific revolution for nanotechnology. Since then, an enormous amount of research has been performed on representative carbon nanomaterials, notably single-wall carbon nanotubes (SWCNTs), discovered in 1993 \cite{IijimaIchihashi93Nature,BethuneetAl93Nature}, and graphene, first isolated in 2004 \cite{NovoselovGeimScience04}.  Not only attracting much attention for future applications in microscale and nanoscale optoelectronic devices, these carbon nanomaterials are stimulating much interest from a fundamental point of view, considered to be one of the most ideal systems that exhibit the quantum mechanical nature of interacting low-dimensional electrons.

While initial investigations on these materials concentrated on DC characteristics, recent theoretical studies have instigated a flurry of new experimental activities to uncover unusual AC properties.   Both SWCNTs and graphene are expected to show exotic terahertz (THz) dynamics that can lead to innovative optoelectronic applications~\cite{PortnoietAl06SPIE,RyzhiietAl06JJAP,KibisetAl07NL,MikhailovZiegler08JPCM,Mikhailov09MJ}.  These properties are inherently related to their unique, low-dimensional band structure, combined with many-body interactions of quantum-confined carriers.


From a fundamental point of view, dynamic (or optical or AC) conductivity $\sigma(\omega)$, where $\omega$ is the (angular) frequency of the applied electric field, is expected to provide a wealth of information on quantum confinement, electron interactions, and disorder. For one-dimensional (1-D) electron systems such as SWCNTs, there have been detailed theoretical calculations of $\sigma(\omega)$, taking into account interactions and disorder to varying degrees (see, e.g., \cite{SablikovShchamkhalova97JETP,RoschAndrei00PRL,Ando02JPSJ,Burke02IEEE,PustilniketAl06PRL} and pp.~219-237 of \cite{Giamarchi04Book}).   For a metallic SWCNT, for example, Ando~\cite{Ando02JPSJ} calculated $\sigma(\omega)$ within a self-consistent Born approximation, which indicated that there can exist non-Drude-like conductivity, depending on the range of scattering potentials.  One particularly important question for metallic SWCNTs is whether the optical conductivity satisfies the so-called $\omega/T$ scaling, i.e.,
the frequency dependence appears only through a particular combination of $\omega$ and $T$, i.e., $\omega/T$~\cite{Giamarchi04Book}, where $T$ is the sample temperature.  This implies that the system is at a quantum critical point~\cite{SietAl01Nature,Sachdev99Book}. 
A Tomonaga-Luttinger liquid has the characteristic feature that it is quantum critical over a finite range of parameters due to enhanced quantum fluctuations in 1-D (see, e.g., \cite{Giamarchi04Book}), and thus, is expected to exhibit a $\omega / T$ scaling law in conductivity.

A number of experimental THz/far-infrared spectroscopic studies have been performed over the last decade on SWCNTs of various forms~\cite{BommelietAl96SSC,UgawaetAl99PRB,ItkisetAl02NL,JeonetAl02APL,JeonetAl04JAP,JeonetAl05JAP,AkimaetAl06AM,BorondicsetAl06PRB,NishimuraetAl07APL,KampfrathetAl08PRL}, producing an array of conflicting results with contradicting interpretations.  This is partly due to the widely differing types of samples used in these studies -- grown by different methods (HiPco, CoMoCAT, CVD, Arc Discharge, and Laser Ablation) and put in a variety of polymer films that are transparent in the THz range.  Nanotubes in most of these samples were bundled and typically consisted of a mixture of semiconducting and metallic nanotubes with a wide distribution of diameters.  Some of the samples used were partially aligned through mechanical stretching, showing some degree of anisotropy in their THz response \cite{JeonetAl02APL,JeonetAl04JAP,AkimaetAl06AM}.  One common spectral feature that many groups have detected is a broad absorption peak around 4~THz (or 135~cm$^{-1}$ or $\sim$17~meV).  This feature, first observed by Ugawa {\it et al}.~\cite{UgawaetAl99PRB}, has been interpreted as interband absorption in mod-3 non-armchair nanotubes with curvature-induced gaps~\cite{UgawaetAl99PRB,BorondicsetAl06PRB,NishimuraetAl07APL,KampfrathetAl08PRL} or absorption due to plasmon oscillations along the tube axis~\cite{AkimaetAl06AM,NakanishiAndo09JPSJ}, but a consensus has not achieved~\cite{SlepyanetAl10PRB}.

\begin{figure}[h]
\begin{center}
  \includegraphics[scale = 0.7]{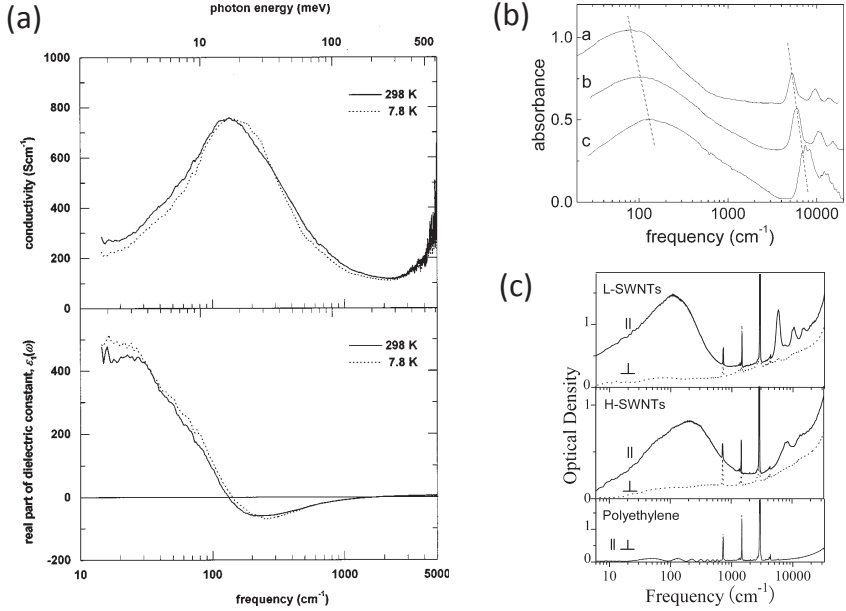}
\end{center}
\caption{Terahertz and infrared spectra reported for different types of SWCNT samples by (a)~Ugawa {\it et al}.~\cite{UgawaetAl99PRB}, (b)~Itkis {\it et al}.~\cite{ItkisetAl02NL}, and (c)~Akima {\it et al}.~\cite{AkimaetAl06AM}.  The universally observed absorption peak around 100~cm$^{-1}$ has different interpretations.}
\label{4THz-peak}
\end{figure}

The AC dynamics of Dirac fermions in graphene have attracted much theoretical attention -- the influence of linear dispersions, two-dimensionality, and disorder has been extensively discussed by many theorists~\cite{ShonAndo98JPSJ,ZhengAndo02PRB,AndoetAl02JPSJ,PeresetAl06PRB,GusyninSharapov06PRB,GusyninetAl06PRL,GusyninetAl07PRL,RyuetAl07PRB,AbergelFalko07PRB,Mishchenko07PRL,GusyninetAl07PRB,RyzhiietAl07JAP,FalkovskyVarlamov07EPJB,MikhailovZiegler07PRL,Mikhailov07EPL,FalkovskyPershoguba07PRB,SheehySchmalian07PRL,HerbutetAl08PRL,KoshinoAndo08PRB,MikhailovZiegler08JPCM,Mishchenko08EPL,Mikhailov09MJ,LewkowiczRosenstein09PRL,RyzhiietAl09APE,Mikhailov09PRB,IngenhovenetAl10PRB,JuricicetAl10PRB}.  However, the influence of electron-electron interactions on the optical conductivity of graphene is somewhat controversial.  Theoretical studies using different methods have led to different conclusions as to the magnitude of many-body corrections to the Drude-like intraband optical conductivity (see, e.g.,~\cite{Mishchenko07PRL,SheehySchmalian07PRL,HerbutetAl08PRL,Mishchenko08EPL} and references cited therein).  Experimentally, while a number of studies have confirmed the so-called universal optical conductivity $\sigma(\omega) = e^2 /4 \hbar$ for {\em interband} transitions in a wide spectra range~\cite{NairetAl08Science,MaketAl08PRL,LietAl08NP}, successful experimental studies of the {\em intraband} conductivity have been very limited~\cite{ChoietAl09APL,HorngetAl11PRB,LiuetAl11JAP,YanetAl11ACS,MaengetAl12NL,Sensale-RodriguezetAl12NC}, except for cyclotron resonance measurements in magnetic fields~\cite{SadowskietAl06PRL,JiangetAl07PRL,DeaconetAl07PRB,HenriksenetAl08PRL,OrlitaetAl08PRL,NeugebaueretAl09PRL,HenriksenetAl10PRL,CrasseeetAl11NP,WitowskietAl10PRB,OrlitaetAl11PRL,CrasseeetAl11PRB,BooshehrietAl12PRB}.  On the other hand, theoretical studies on {\em nonlinear and non-equilibrium properties} of graphene have emerged in recent years, urging further experiments to be done using time-resolved and/or nonlinear optical spectroscopy.  For example, Mikhailov and Ziegler~\cite{MikhailovZiegler08JPCM,Mikhailov07EPL}, using a semiclassical approximation, have shown that  Dirac fermions in graphene with a dispersion $\varepsilon$($\emph{\textbf{p}}$) = $v_{\rm F}$$|$$\emph{\textbf{p}}$$|$ (where $v_{\rm F}$ is the Fermi velocity) in the presence of an applied AC electric field $E_{x}$($t$) = $E_{0}$$\cos$$\omega t$ would produce an AC current $j_x(t) =  e n_s v_{\rm F} (4 / \pi ) ( \sin\omega t + \frac{1}{3}\sin3\omega t + \frac{1}{5}\sin5\omega t + \cdots )$ {\em within linear response}.  Note that not only does the current oscillate at the frequency of the applied field $\omega$, but it also contains all the odd harmonics.   This should be contrasted to conventional semiconductors such as GaAs, which, within linear response, will only produce an AC current with frequency $\omega$, i.e., $j_x(t) = \frac{n_s e^2 E_0}{m^* \omega}\sin\omega t$, where $m^*$ is the effective mass.  Therefore, the dynamics of graphene in an AC electric field are intrinsically nonlinear, and efficient frequency multiplications for THz generation can be expected for microwave-driven graphene.

\begin{figure}
\begin{center}
\includegraphics[scale = 0.45]{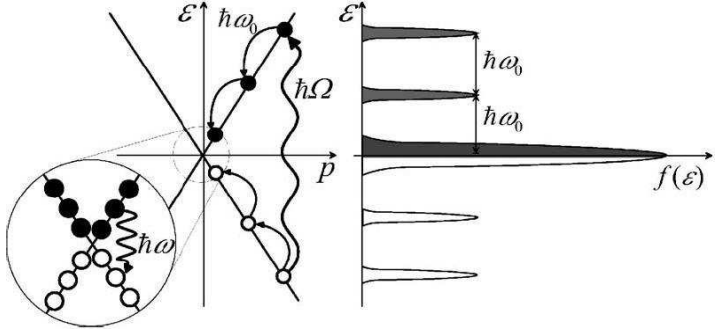}
\end{center}
\caption{Sketch of graphene bandstructure and energy distributions of photogenerated electrons and holes under interband optical pumping of photon energy $\hbar$$\Omega$. A cascade of phonon emissions ($\hbar$$\omega$$_{0}$) is followed by population inversion and gain for radiation $\hbar$$\omega$ for sufficiently strong optical pumping. Reproduced from Ref.~\cite{RyzhiietAl07JAP}}
\label{RyzhiiJAP} 
\end{figure}

Another outstanding theoretical prediction is THz amplification in optically pumped graphene~\cite{RyzhiietAl07JAP}.  Ryzhii and co-workers calculated the dynamic conductivity of a non-equilibrium 2D electron-hole system in graphene under interband optical excitation. Both interband and intraband transitions were taken into account in their model.  Under optical pumping with photon energy $\hbar$$\Omega$, electrons and holes are photogenerated with energy $\varepsilon_0$ = $\hbar \Omega$/2 and emit a cascade of optical phonons with energy $\hbar \omega_{0}$, leading to population accumulation of the bottom of the conduction band (by electrons) and the top of the valence band (by holes), as shown in Fig.~\ref{RyzhiiJAP}. They demonstrated that sufficiently strong optical pumping will result in population inversion, making the real part of the net AC conductivity negative, i.e., amplification. Due to the gapless energy spectrum, this negative AC conductivity takes place in the range of THz frequencies. This effect might be used in graphene-based coherent sources for THz radiation.


\section{Anisotropic THz Conductivity of Macroscopically Aligned Single-Wall Carbon Nanotubes}
\label{sec:SWNTsCond}

\subsection{Introduction}
The 1-D character of confined carriers in SWCNTs manifests itself as strong anisotropy in electric, magnetic, and optical properties~\cite{DresselhausetAl01Book,Oconnell06Book,JorioetAl08Book}. Individual metallic SWCNTs have been shown to be excellent 1-D electrical conductors with very long coherence lengths \cite{DekkerNature97}, while individual semiconducting SWCNTs have been shown to absorb and emit light only when the light polarization is parallel to the tube axis \cite{SrivastavaetAl08PRL}. Moreover, individualized SWCNTs, both metallic and semiconducting, have been shown to align well in an external magnetic field \cite{IslametAl04PRL,ZaricetAl04Science,ZaricetAl04NL,IslametAl05PRB,TorrensetAl07JACS,ShaveretAl09ACS} due to their anisotropic magnetic susceptibilities \cite{AjikiAndo93BJPSJ,Lu95PRL,MarquesetAl06PRB}, and aligned SWCNTs exhibit strong linear dichroism due to their anisotropic optical properties. Even in bundled samples where SWCNTs form aggregates through van der Waals attraction, their anisotropic properties are expected to be preserved to a large degree, even though systematic optical spectroscopy studies have been limited \cite{MurakamietAl05PRL,FaganetAl07PRL,ShaveretAl08PRB} due to the rarity of ensemble samples in which the SWCNTs are highly aligned.

Here, we describe results of polarization-dependent THz transmission measurements on films of macroscopically-aligned SWCNTs, which demonstrate an extremely high degree of anisotropy~\cite{RenetAl09NL,RenetAl12NL}.  Strikingly, when the THz polarization is perpendicular to the nanotube alignment direction, practically no absorption is observed despite the macroscopic thickness of the film.  On the other hand, when the polarization is parallel to the alignment direction, there is strong absorption. The degree of polarization in terms of absorbance is 1 and the reduced linear dichroism is 3, throughout the entire frequency range of our experiment (0.1-2.2 THz). Using the theory of linear dichroism for an ensemble of anisotropic molecules~\cite{RodgerNorden97Book}, we show that this value of reduced dichroism (i.e., 3) is possible only when the nematic order parameter ($S$) is 1. These observations are a direct result of the 1-D nature of conduction electrons in the nanotubes and at the same time demonstrate that any misalignment of nanotubes in the film must have characteristic length scales much smaller than the wavelengths used in these experiments (1.5~mm to 150~$\mu$m). 

These results on aligned SWCNTs, as well as a recent report by Kyoung {\it et al}.~on aligned multi-wall CNTs~\cite{KyoungetAl11NL}, indeed suggest that aligned CNT films perform as ideal linear polarizers in the THz frequency range.  These aligned films (or stacks of films~\cite{RenetAl12NL}) exhibit comparable performance to commercial wire-grid technology but have added benefits of (i) broadband THz absorption driven by the inherent 1-D character of the CNTs and (ii) mechanical robustness in diverse operation conditions. Namely, in comparison to wire-grid technology, the THz performance of these materials is driven not by the precise structure of the conductive wires but rather the inherent anisotropic THz absorption properties of aligned CNTs.

\subsection{Experimental Methods}

The aligned SWCNT films studied were produced by natural self-assembly of SWCNTs into densely packed and highly aligned macroscopic materials during synthesis. Utilizing optical lithography to define the pitch between lines of catalyst, water-assisted chemical vapor deposition was employed to grow aligned SWCNTs in high aspect-ratio lines.  Following growth, the lines were transferred to a sapphire or silicon substrate to produce a THz polarizer~\cite{RenetAl09NL,RenetAl12NL,PintetAl10ACS}.  To optimize the device performance with respect to extinction, multiple layers were stacked until full extinction of linearly polarized THz radiation was achieved in a configuration where the THz field is parallel to the alignment. The benefit of this approach compared to other possible approaches is the simplicity of the contact-transfer process that makes this scalable to large areas. There are no known limitations in this approach that would inhibit scaling to full wafers or even continuous roll-to-roll growth and contact transfer techniques to yield high-throughput production of THz devices.

The THz-TDS system used a Ti:Sapphire femtosecond laser that excited either a ZnTe crystal or a photoconductive antenna to generate and detect coherent THz radiation.  The sample was positioned at the THz focus and was rotated about the propagation direction of the THz beam, which changed the angle, $\theta$, between the nanotube axis and the THz polarization direction from 0$^{\circ}$ to 90$^{\circ}$, as shown in Fig~\ref{SWNTsTHzPolarizer}.   More details about the experimental setup are described elsewhere~\cite{RenetAl09NL,RenetAl12NL}.
With an intrinsic silicon substrate, which is transparent in almost the entire infrared (IR) range, we were also able to  perform similar polarization-dependent transmission measurements using Fourier transform infrared (FTIR) spectroscopy.

\subsection{Experimental Results}

\begin{figure}
\begin{center}
  \includegraphics[scale = 0.6]{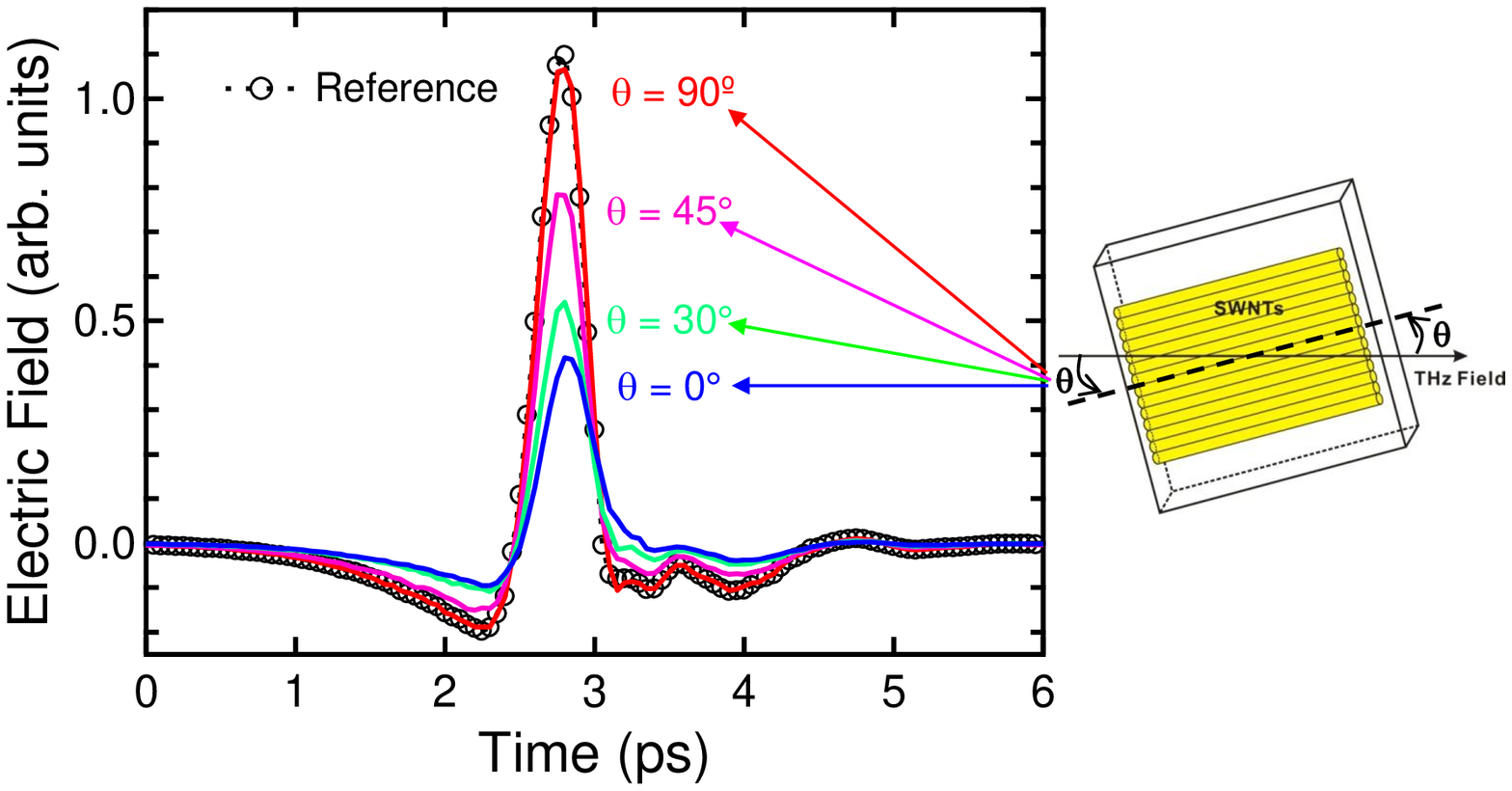}
\end{center}
\caption{Transmitted THz electric field signals for the nanotube film for different polarization angles $\theta$ (colored solid curves) and for the reference sapphire substrate (dashed curve with open circles).  Adapted from \cite{RenetAl09NL}.}
\label{SWNTsTHzPolarizer}
\end{figure}

Figure~\ref{SWNTsTHzPolarizer} shows transmitted time-domain waveforms for four different angles ($\theta$ = 0$^{\circ}$, 30$^{\circ}$, 45$^{\circ}$, and 90$^{\circ}$) between the THz polarization direction and the CNT alignment direction, together with the waveform transmitted through a reference sapphire substrate with no nanotubes.  Note that the 90-degree trace precisely coincides with the reference waveform, which means that there is no attenuation for this THz polarization (perpendicular to the tubes).  From the measured absorption anisotropy, we calculated the reduced linear dichrosim to be 3, corresponding to a nematic order parameter of 1~\cite{RenetAl09NL}.   This result suggests that any misalignment of nanotubes in the film is negligible compared to the wavelengths used in these experiments (1.5~mm $-$ 150~$\mu$m).

\begin{figure}
\begin{center}
\includegraphics[scale = 0.45]{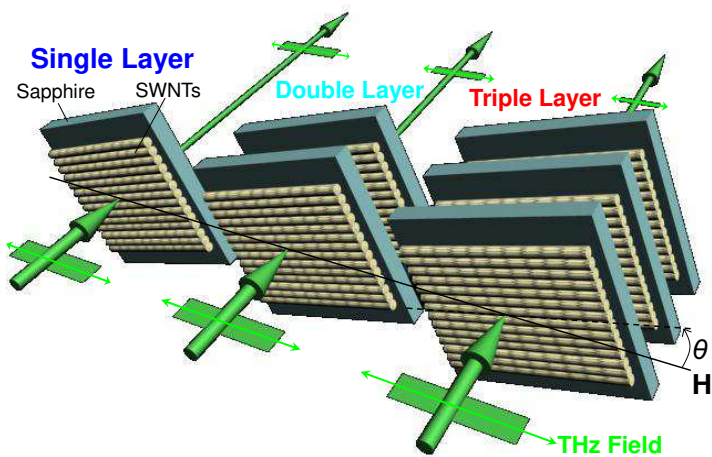}
\end{center}
\caption{Scheme showing the use of multiple SWCNTs films to produce high performance polarizers, as discussed in the text. Adapted from \cite{RenetAl12NL}.}
\label{SWNTsFilmStacks}
\end{figure}

\begin{figure}
\begin{center}
  \includegraphics[scale = 0.6]{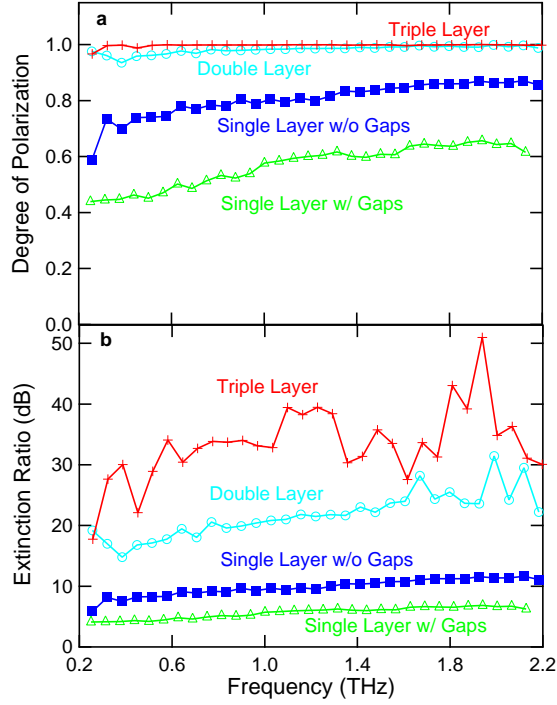}
\end{center}
\caption{Polarizer parameters for the SWNTs polarizers: (a)~Degree of polarization and (b)~extinction ratio of the THz polarizers with different thicknesses as a function of frequency in the 0.2$-$2.2 THz range.  For the optimized triple-layer SWCNT polarizer, the averaged value of the extinction ratio in this frequency range is 33.4 dB.  Adapted from \cite{RenetAl12NL}.}
\label{SWNTsStacksDOPER} 
\end{figure}

In order to optimize the device performance with respect to extinction, multiple layers were stacked in parallel with each other, as shown in Fig.~\ref{SWNTsFilmStacks}, until full extinction of linearly polarized THz radiation was achieved in a configuration where the THz field is parallel to the alignment.  To elucidate polarizer performance, we calculated the degree of polarization (DOP), defined as DOP = ($T_{\perp} - T_{\parallel}$)/($T_{\perp} + T_{\parallel}$), and the extinction ratio (ER), defined as ER = $T_{\parallel}$/$T_{\perp}$ , where $T_{\parallel}$ is the transmittance for the parallel case, and $T_{\perp}$ is the transmittance for the perpendicular case.  As shown in Fig.~\ref{SWNTsStacksDOPER}(a), the DOP value of our SWCNTs polarizer increases with the SWCNTs film thickness and reaches a value at 99.9$\%$ throughout the whole measured frequency range for a triple-stacked film.  Figure~\ref{SWNTsStacksDOPER}(b) indicates that the ER value is increasing dramatically with the film thickness, achieving an average value of 33.4~dB in the 0.2$-$2.2~THz range --- two orders of magnitude better than the thinner SWCNT films.  These results demonstrate the remarkable utility of aligned SWCNTs for THz applications.

\begin{figure}[h]
\begin{center}
  \includegraphics[scale = 0.6]{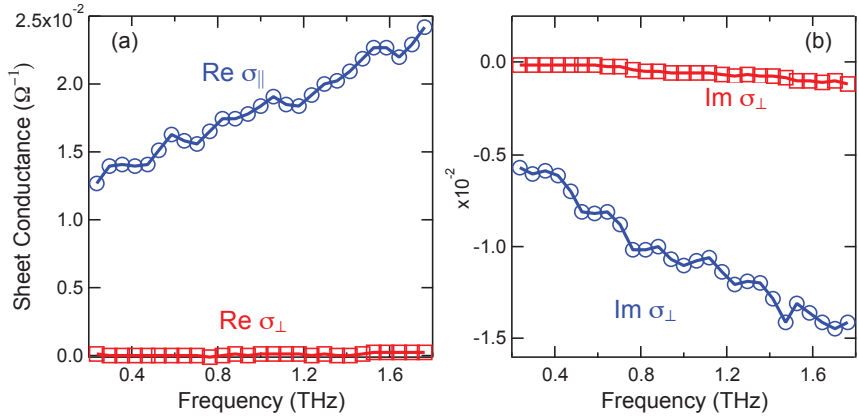}
\end{center}
\caption{(a) Real and (b) imaginary parts of the parallel and perpendicular elements of the dynamic conductivity tensor for the aligned SWCNT film extracted from the THz time-domain signals.}
\label{CondTensor}
\end{figure}

We Fourier-transformed the time-domain data of the single highly aligned SWCNT thin film shown in Fig.~\ref{SWNTsTHzPolarizer} into the frequency domain and extracted the parallel and perpendicular elements of the the dynamic complex conductivity tensor $\tilde\sigma(\omega)$, i.e., $\sigma_{\parallel}(\omega)$ and $\sigma_{\perp}(\omega)$, respectively, as shown in Fig.~\ref{CondTensor}.  Highly anisotropic responses are again observed.  At 90$^{\circ}$ (THz polarization perpendicular to the nanotube axis), the real part of the conductivity is zero throughout the entire frequency range, showing no sign of absorption.  In contrast, at 0$^{\circ}$ (THz polarization parallel to the nanotube axis) the conductivity is finite and reaches $\sim$140~S$\cdot$cm$^{-1}$ at 1.8~THz.  Note that the real part monotonically increases with increasing frequency in this frequency range.  This is consistent with the existence of a peak at a higher frequency, as observed by many other groups on various types of SWCNTs, as discussed previously and shown in Fig.~\ref{4THz-peak}.  The results shown here unambiguously demonstrate that such a finite-frequency peak appears only in the parallel component of the dynamic conductivity tensor.


\section{THz and Infrared Spectroscopy of Gated Large-Area Graphene}
\label{sec:THzGr}

Graphene has been intensively studied since its first isolation in 2004~\cite{NovoselovGeimScience04,GeimNovoselovNMater07,NovoselovGeimNature05}. This zero-gap semiconductor consisting of a single layer of $sp^{2}$-bonded carbon atoms arranged into a 2-D honeycomb lattice possesses a photon-like, linear energy dispersion, which is expected to lead to exceptionally nonlinear electro-dynamic properties~\cite{Mikhailov07EPL,RyzhiietAl07JAP}.  Since both intraband and interband transitions in graphene are expected to be sensitive to the location of the Fermi energy, AC studies with tunable carrier concentration would provide significant new insights into the dynamics of 2-D Dirac fermions.

The Fermi level can be tuned either by doping or gating.  Substitution of carbon atoms in graphene by nitrogen and boron have been attempted, but this dramatically decreases the mobility by breaking its lattice structure.  Physically adsorbed molecules can also dope graphene, but this is not stable and leads to a suppression of mobility as well.  So far, applying a controllable gate voltage to graphene to transfer carriers from the doped silicon substrate is still the most common and reliable way.  By utilizing applied gate voltages, different groups have observed tunable interband optical transitions~\cite{LietAl08NP,WangetAl08Science}, tunable intraband far-infrared conductivity~\cite{HorngetAl11PRB}, and a systematic G-band change with gate voltage in Raman spectra~\cite{PisanaetAl07NM,YanetAl07PRL}.

Here, we describe our THz and IR study of {\em large-area} (centimeter scale) graphene with an electrically tunable Fermi level ($E_f$).  In a field effect transistor consisting of graphene on a SiO$_2$/$p$-Si substrate, the intensity of the transmitted THz electromagnetic wave was observed to change with the gate voltage. Both the Drude-like intraband conductivities and the `2$E_f$ onset' of the interband transitions, monitored through THz time-domain spectroscpy and Fourier-transform infrared spectroscopy, respectively, were both modulated by the gate voltage.

\begin{figure}[h]
\begin{center}
\includegraphics[scale = 0.85]{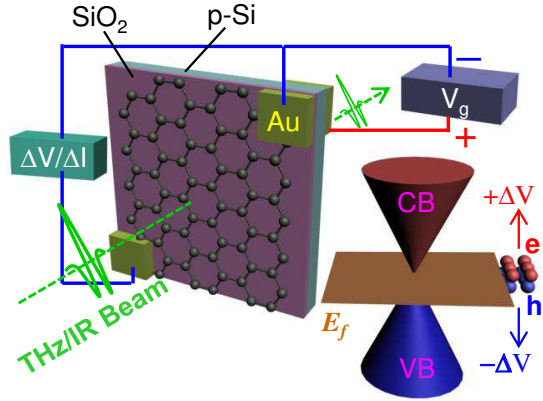}
\caption{Experimental configuration sketch of gate-voltage-dependent THz-TDS/FTIR transmission measurements on graphene/SiO$_{2}$/$p$-Si sample.  Adapted from \cite{RenetAl12NL2}.}
\end{center}
\label{GateExpConf} 
\end{figure}

The large-area graphene sample used for this spectroscopy study was grown from a solid state carbon source -- poly(methyl methacrylate) (PMMA)~\cite{SunetAl10Nature}.  Grown single-layer graphene was transferred to a $\sim$1.5~cm $\times$ 1.5~cm $p$-type silicon wafer (5-10~${\rm \Omega}$-cm) with a 300-nm thick SiO$_2$ layer.  Gold electrodes were then deposited on corners of the 8~mm $\times$ 8~mm graphene film and on the back of the p-Si substrate (see Fig.~\ref{GateExpConf}.  The THz wave was normal incident onto the center of graphene, and the transmitted THz wave for was detected and analyzed as a function of gate voltage.

\begin{figure}[h]
\begin{center}
\includegraphics[scale = 0.55]{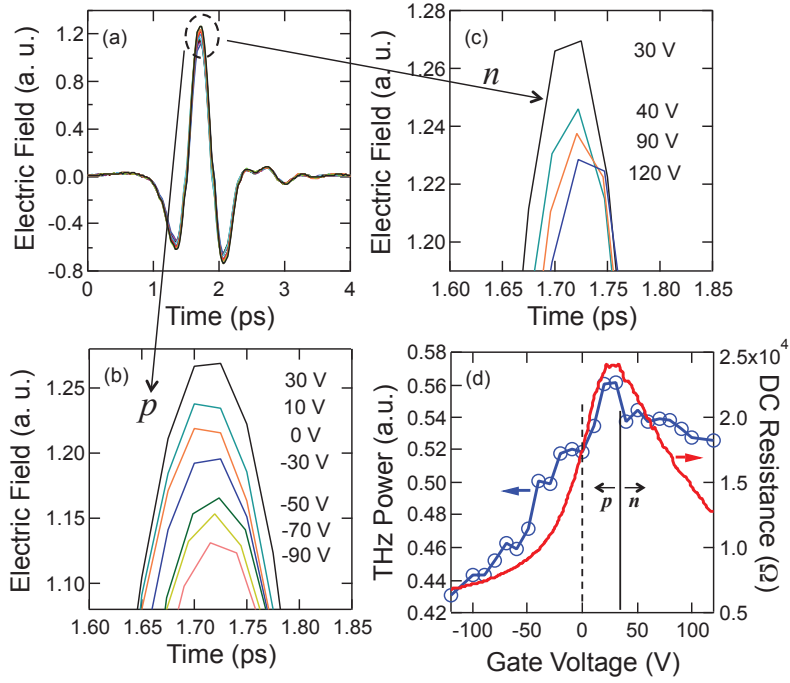}
\end{center}
\caption{Gate-voltage-dependent THz-TDS/FTIR transmission measurements on graphene/SiO$_{2}$/$p$-Si sample: (a) Transmitted THz waveforms under different applied gate voltages. (b) Monotonic change of THz signals with the gate voltage in the hole regime. (c) Monotonic change of THz signals with the gate voltage in the electron regime. (d) Transmitted THz power (blue line with open circles) and DC resistance of graphene (red solid line) as a function of gate voltage.}
\label{THzTDSGrGating}
\end{figure}

Figure \ref{THzTDSGrGating}(a) shows gate-voltage-dependent transmitted THz waveforms, the signal peaks of which are zoomed in and illustrated in Fig.~\ref{THzTDSGrGating}(b) and (c).  At +30~V, the Fermi energy is at the charge neutrality point (CNP), and thus, the highest THz transmission is obtained.  At all other voltages above and below +30~V, THz transmission decreases (or absorption increases) monotonically with the voltage change, as shown Fig.~\ref{THzTDSGrGating}(b) and (c).  Figure \ref{THzTDSGrGating}(d) shows the transmitted THz beam power as a function of gate voltage (blue circled line), demonstrating that +30~V is indeed closest to the CNP and the unbiased (0-V) point is on the $p$-doped side.  The DC resistance of the device was measured {\it in situ} and is plotted as a function of gate voltage in Fig.~\ref{THzTDSGrGating}(d) (the red trace), showing agreement with the gate dependence of the transmitted THz power.

\begin{figure}[h]
\begin{center}
\includegraphics[scale = 0.6]{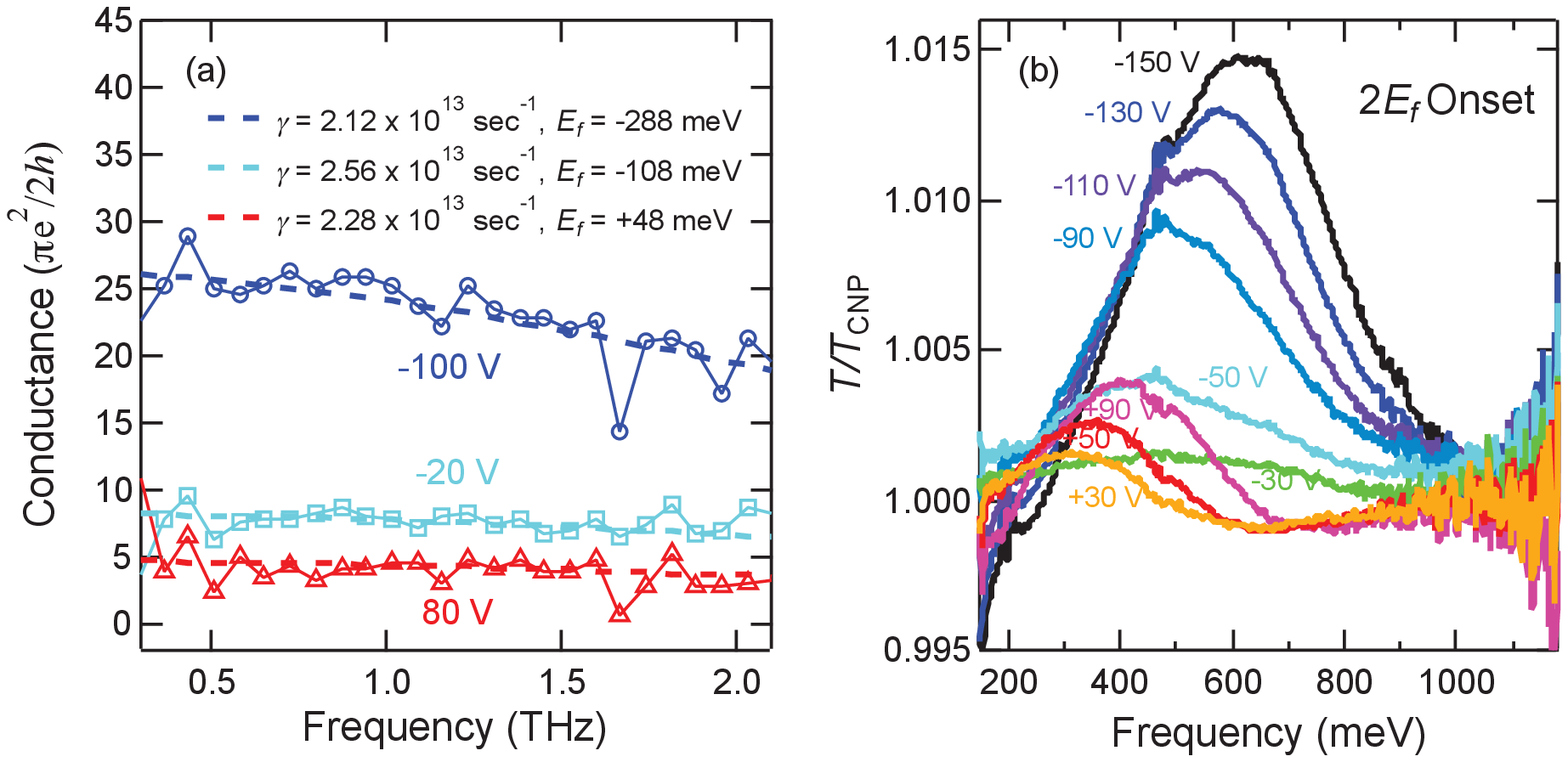}
\end{center}
\caption{Dynamic conductivity of the large-area graphene in THz/IR frequency range under various gate voltages. (a)~Real and part of the 2D conductance for intraband conductivity in the THz range. (b)~2$E_f$ onset in the MIR transmission spectra for interband conductivity.}
\label{GrapheneTHz-IR}
\end{figure}

After Fourier transforming the time-domain data shown in Fig.~\ref{THzTDSGrGating}, we obtained the corresponding transmission spectrum in the frequency domain with the signal taken at +30~V (the CNP) as a reference.  Using standard analysis methods appropriate for thin conducting films, the 2D THz conductance of the graphene sample was extracted as a function of frequency, as shown for three representative gate voltage values in Fig.~\ref{GrapheneTHz-IR}(a).  The AC conductance is seen to decrease with increasing frequency throughout the range of 0.3-2.1~THz (10-70 cm$^{-1}$), in contrast to the case of SWCNTs (see Fig.~\ref{CondTensor}).  The THz data is combined with FTIR data taken at higher frequencies (100-600~cm$^{-1}$).  

Also shown in Fig.~\ref{GrapheneTHz-IR}(a) are theoretical fits to the data (colored dashed lines) to deduce the Fermi energy ($E_f$) and scattering rate ($\gamma$), using~\cite{Mikhailov09MJ}
\begin{equation}
\sigma_{\rm intra} (\omega) = {2i e^2 k_B T \over \pi \hbar^2 (\omega + i \gamma)} \ln(e^{E_f/ k_{\rm B}T} + e^{-E_f/ k_{\rm B}T})
\end{equation}
where $e$ is the electronic charge, $\hbar$ is the reduced Planck constant, $k_{\rm B}$ is the Boltzmann constant, and $T$ is the temperature (= 300~K in our experiments).  The larger the gate voltage is , measured from the CNP, the larger the Fermi energy, and as a result, the larger the overall conductance.  The extracted values for the Fermi energy and scattering rate are listed within Fig.~\ref{GrapheneTHz-IR}(a).  The obtained values for $\gamma$ are on the order of 2 $\times$ 10$^{13}$ sec$^{-1}$ (or a scattering time of $\sim$50~fs), which agrees with our results on the same type of graphene samples obtained from high-field mid-IR cyclotron resonance measurements~\cite{BooshehrietAl12PRB}.   The obtained $E_f$ values are also generally consistent with the ``2$E_f$ onset'' observed in the MIR range, shown in Fig.~\ref{GrapheneTHz-IR}(b).  Here, the ratio of the transmission spectrum under a certain gate voltage by that under the CNP gate voltage is plotted for various gate voltages.  As we can see, as the gate voltage was changed, the 2$E_f$ onset peak shifts monotonically, indicating a tuned Fermi level.  These gated large-area graphene samples are thus promising for further basic studies of low-frequency phenomena with a tunable Fermi energy as well as for manipulation of THz and infrared waves.


\section{Summary}
\label{sec:conclusion}
In this article, we reviewed our recent work progress on the optical conductivities of low-dimensional carriers in single-wall carbon nanotubes (SWCNTs) and graphene, two types of carbon nanomaterials with molecules structured with $sp^{2}$-bonded carbon atoms.  Polarization-dependent THz transmission measurements on highly aligned SWCNT films revealed extreme anisotropy.  Based on this, we synthesized polarizers made of aligned SWCNTs on sapphire substrates that exhibit remarkable broadband performance in the THz frequency range, with degrees of polarization of 99.9$\%$ from 0.2 to 2~THz, and extinction ratios of up to $\sim$35~dB.  This material yields broadband polarization and extinction features that outperform the conventional wire-grid polarizers. Through a proper model, the dynamic complex conductivity tensor elements of these SWCNT films were determined from the THz-TDS data, and both the real part and the imaginary part of the parallel conductivity show a non-Drude frequency dependence.  We also measured the transmission spectra of large-area graphene films grown from solid state carbon source on SiO$_{2}$/$p$-Si through THz-TDS and Fourier transform infrared spectroscopy, and determined the optical conductivity of graphene in the THz and infrared ranges.  The Fermi energy was tuned by applying a gate voltage to the device, which, in turn, modulated the transmission of THz and IR waves.  The frequency dependence of the transmission allowed us to determine the Fermi energy and scattering time at each gate voltage, both from the THz (intraband) and IR (interband) spectra.

\begin{acknowledgements}
This work was supported by the National Science Foundation (through Grant No.~OISE-0530220 and No.~OISE-0968405), the Department of Energy (through Grant No.~DEFG02-06ER46308), and the Robert A.~Welch Foundation (through Grant No.~C-1509).  We thank Cary L.~Pint and Robert H.~Hauge for providing us with the aligned carbon nanotube films and Jun Yao, Zhengzong Sun, Zheng Yan, Zhong Jin, and James M.~Tour for providing us with the large-area graphene samples.
\end{acknowledgements}

\pagebreak


\end{document}